\documentstyle[sprocl]{article}

\catcode`\@=11
\def\@cite#1#2{{\if@cghi$\!^{#1}$\else$[{#1}]$\fi\if@tempswa\typeout
        {IJCGA warning: optional citation argument
        ignored: `#2'} \fi}}
\catcode`\@=\active

\def\wt{\widetilde}

\def\sst{\scriptscriptstyle}
\def\ra{\rightarrow}

\def\cB{\cal B}

\def\cD{\cal D}

\def\cG{\cal G}

\def\cL{\cal L}

\def\cP{\cal P}

\def\tH{{\tilde H}}
\def\tv{{\tilde v}}
\def\tO{{\tilde \Omega}}
\def\wt{\widetilde}

\def\p{\partial}

\def\be{\begin{equation}}\def\ee{\end{equation}}
\def\ba{\begin{array}}\def\ea{\end{array}}
\def\bea{\begin{eqnarray}}\def\eea{\end{eqnarray}}
\def\bd{\begin{document}}\def\ed{\end{document}}
\def\fin{\end{thebibliography}\end{document}}

\let\la=\label

\let\bm=\bibitem
\def\nn{\nonumber}
\def\qq{\quad\quad}
\let\fr=\frac

\def\ft#1#2{{\textstyle{{\scriptstyle #1}\over {\scriptstyle #2}}}}
\def\fft#1#2{{#1 \over #2}}
\def\sst#1{{\scriptscriptstyle #1}}
\def\oneone{\rlap 1\mkern4mu{\rm l}}
\newcommand{\eq}[1]{(\ref{#1})}
\newcommand{\w}[1]{\\[0.#1cm]}

\def\Dot#1{\buildrel{_{_{\hskip 0.01in}{\normalsize\bullet}}}\over{#1}}
\def\Hat#1{\widehat{#1}}
\def\bsh{\backslash}
\def\dA{\Dot A}
\def\dB{\Dot B}
\def\dC{\Dot C}
\def\a{\alpha}
\def\b{\beta}
\def\c{\gamma}\def\C{\Gamma}\def\cdt{\dot\gamma}
\def\d{\delta}\def\D{\Delta}\def\ddt{\dot\delta}
\def\e{\epsilon}\def\vare{\varepsilon}
\def\f{\phi}\def\F{\Phi}\def\vf{\varphi}
\def\h{\eta}
\def\k{\kappa}
\def\l{\lambda}\def\L{\Lambda}
\def\m{\mu}
\def\n{\nu}
\def\r{\rho}
\def\s{\sigma}\def\S{\Sigma}
\def\t{\tau}
\def\th{\theta}\def\Th{\Theta}\def\tb{\bar\theta}
\def\x{\xi}
\def\o{\omega}\def\O{\Omega}

\def\ua{\underline{\a}}
\def\ub{\underline{\phantom{\a}}\!\!\!\b}
\def\uc{\underline{\phantom{\a}}\!\!\!\c}
\def\um{\underline{\mu}}
\def\ud{\underline\d}
\def\ue{\underline\e}
\def\una{\underline a}\def\unA{\underline A}
\def\unb{\underline b}\def\unB{\underline B}
\def\unc{\underline c}\def\unC{\underline C}
\def\und{\underline d}\def\unD{\underline D}
\def\une{\underline e}\def\unE{\underline E}
\def\unf{\underline{\phantom{e}}\!\!\!\! f}\def\unF{\underline F}
\def\ung{\underline g}
\def\unm{\underline m}\def\unM{\underline M}
\def\unn{\underline n}\def\unN{\underline N}
\def\unp{\underline{\phantom{a}}\!\!\! p}\def\unP{\underline P}
\def\unH{\underline{H}}
\def\unF{\underline{F}}\def\unT{\underline{T}}
\def\ovA{\overline{A}}\def\ovB{\overline{B}}
\def\uC{{\underline C}}

\def\smpl{ +}
\def\smm{ -}
\def\ns{\normalsize}

\def\t1{
\begin{table}
{\footnotesize
\begin{tabular}{|c|c|c|c|c|}
\hline
&&&&\\
$D$ & $N$ & Scalar Manifold $G/H$    & Gauge Group $K \subseteq G$
& Matter Sector \\
&&&&\\
\hline
  &&&&\\
  10 & (2,0) & $SU(1,1)/U(1)$ & ---- & ---- \\
&&&&\\
  \hline
  &&&&\\
  9 & 2 & $GL(2,R)/SO(2)$ & $SO(2)$ & ---- \\
  &&&&\\
  &  1  &  $SO(n,1)/SO(n)$  &  ${\rm dim\ K\subseteq n+1}$  &  $n$\ Maxwell \\
  &&&&\\
  \hline
  &&&&\\
  8 & 2 & $SL(3,R)/SO(3) \times SL(2,R)/SO(2)$ & $SO(3)$ & ---- \\
  &&&&\\
  &1& $SO(n,2)/SO(n)\times SO(2)$ & ${\rm dim\ K\subseteq n+2}$ & $n$\ Maxwell\\
  &&&&\\
  \hline
  &&&&\\
   7 & 2 & $SL(5,R)/SO(5) $ & $SO(5)$ & ---- \\
  &&&&\\
  & 1 & $SO(n,3)/SO(n)\times SO(3)$ & ${\rm dim\ K\subseteq n+3}$&$n$\ Maxwell\\
  &&&&\\
  \hline
  &&&&\\
  6 & (2,2) & $SO(5,5)/SO(5)\times SO(5)$ & $SO(5)$ & ---- \\
  &&&&\\
      &  (2,0) &  $SO(n,5)/SO(n)\times SO(5)$  &  ---   &   $n$\ Tensor \\
  &&&&\\
  &(1,1)&$SO(n,4)/SO(n)\times SO(4)$&${\rm dim\ K\subseteq n+4}$ &$n$\ Maxwell\\
  &&&&\\
      &  (1,0) &  Quaternionic Kahler  & $Sp(1)\times K'$ &   $n$\ Hyper  \\
  &&&&\\
      &        &  $SO(n,1)/SO(n)$  &  ----  &   $n$\ Tensor  \\
  &&&&\\
  \hline
  &&&&\\
  5 & 4 & $E_6/USp(8)$ & $ SO(6) $ & ---- \\
  &&&&\\
      &  3  &  $SU^{*}(6)/USp(6)$ &   $SU(3)\times U(1)$ & ---- \\
  &&&&\\
  & 2 & $SO(n,5)/SO(n)\times SO(5)$&${\rm dim\ K\subseteq n+5}$ & $n$\ Maxwell\\
  &&&&\\
      &        & Quaternionic Kahler   & $Sp(1)\times K'$ &  $n$\ Hyper \\
  &&&&\\
  &1&
  $SO(n-1,1)\times SO(1,1)/SO(n-1)$ &${\rm dim\ K\subseteq n}$ &$n$\ Maxwell\\
  &&&&\\
      &        &  $E_{6(-26)}/F_4$  & $SU(3)$ &   25 \ Maxwell      \\
  &&&&\\
        &      &  $SU^*(6)/Sp(3)$  &$SU(3)$ &  13 \ Maxwell      \\
  &&&&\\
        &      &  $SL(3,C)/SU(3)$  & $SU(3)$ &   7 \ Maxwell      \\
  &&&&\\
        &      &  $SL(3,R)/SO(3)$  & $SO(2)$ &   4 \ Maxwell      \\
  &&&&\\
  \hline
  \end{tabular}}
  \caption{ \small Supergravities in $D >4 $ dimensions with $N$
  supersymmetry and nontrivial sigma model sectors. }
  \end{table}
  }

\begin{document}

\title{PROPERTIES OF GAUGED SIGMA MODELS \footnote{Contribution to
the {\it R. Arnowitt Fest: A Symposium on Supersymmetry and
Gravitation}, College Station, TX, 5-7 April 1998. }}

\author{ R. PERCACCI}

\address{International School for Advanced Studies, Trieste, Italy\\
Istituto Nazionale di Fisica Nucleare, Sezione di Trieste}

\author{ E. SEZGIN}

\address{Center for Theoretical Physics, Texas A\&M University,
College Station, \\ TX 77843, USA}

\maketitle \abstracts {Nonlinear sigma models arise in supergravity
theories with or without matter couplings in various dimensions
and they are important in understanding the duality symmetries of
$M$ theory. With this motivation in mind, we review  the
salient features of gauged and ungauged nonlinear sigma models
with or without Wess-Zumino terms for general target spaces in a
minimal as well as lifted formulation. Relevant to the question of
finding interesting vacua of gauged supergravity theories is the
highly constrained potential which arises naturally in these
theories. Motivated by this fact, we derive a general and simple
formula for a gauge invariant potential of this kind.}


\section{Introduction}


Nonlinear sigma models arise naturally in globally or locally
supersymmetric field theories. The scalar fields which parametrize
the sigma model manifold either arise from matter multiplets or
they are part of a supergravity multiplet. In either case, once
coupled to supergravity, scalar fields always seem to form a sigma
model manifold, thereby lending themselves to a geometrical
treatment. It is this geometrical aspect, together with the
attendant symmetries of the system, which makes it possible to
control what otherwise might be very complicated and nonlinear
structure of the couplings of scalar fields to supergravity.

It is important to understand the structure of the supergravity
theories in presence of scalar field couplings in all possible
dimensions since the global and local symmetries which govern
their structure \cite{cj} turn out to be very powerful in
capturing various important properties of a deeper underlying
theory, such as M-theory. For example, the hidden symmetries of
supergravity theory, which are intimately related to the structure
of the sigma model sectors involved, shed light on the all
important duality symmetries of M-theory.

It is impossible to list all possible sigma model manifolds that
can arise in supergravity theories. However, it is useful to give
a list of a large class of such manifolds, at least to give a
flavor of what is involved. In dimensions $D>4$ the sigma model
manifolds that are known to arise form almost a manageable list.
To begin with they are primarily $G/H$ coset spaces of one kind or
another. Typically a subgroup $K$ of the group $G$ can be and has
been gauged. A probably incomplete but rather extensive list of
coset spaces and gauge groups involved in $D>4$ supergravities is
provided in Table 1, in the Appendix. The list gets more involved
in $ D \le 4$ and we will not attempt to construct it here. Let us
emphasize, however, that not in all $D\le 4$ are the sigma model
manifolds necessarily coset manifolds. For example, in $N=1$
supergravity coupled to scalar multiplets in $D=2$, it is well
known that the scalar manifold can be an arbitrary riemannian
manifold. What makes it difficult to review thoroughly all
possible sigma models that can arise in various supersymmetric
field theories is that supersymmetry imposes elaborate set of
geometrical conditions on the geometry of the sigma model
manifolds depending on the dimensions and number of
supersymmetries involved. These conditions need to be analyzed
case by case and they often require the existence of certain
structures on the sigma model manifolds, the most typical of which
are the complex structures. One pattern is clear though: as the
number of supersymmetries increases, the constraints on the sigma
model manifold become more and more stringent.

Most of the supergravity theories and their gaugings listed in
Table 1 have already been constructed. For a detailed review, we
refer the reader to \citelow{ss}. Some are still to be
constructed. For example, gauged $D=9$ supergravities and (gauged)
supergravities coupled to $n$ Maxwell fields in
$N=(1,1), D=6$ and $N=2, D=5$ seem not to have been constructed, but
the expected symmetries are nonetheless listed in Table 1. Gauging
of the sigma models associated with the scalars of tensor
multiplets do not seem to be possible because one does not know
how to construct gauge covariant field strengths for tensor fields
which are in non-singlet representations of the gauge group.

Sigma model manifolds consisting of a real line ${\bf R}$ are not
listed in Table 1. The pure $(1,1)$ supergravity in $D=6$, for
example has one scalar field. This theory also contains six vector
fields of which four can be used to gauge an $Sp(1)\times U(1)$
subgroup of the automorphism group. This example shows that
gauging automorphism groups of a supergravity theory does not
necessarily involve nonlinear sigma model sectors. Other examples
of this phenomenon arise in $N=1,2$ supergravities in $D=5$. What
is also not listed in Table 1 is the noncompact versions of $K$
that can be gauged. The significance of the gauge group $K'$ shown
in Table 1 will be explained in section 9, when we consider the
potential in gauged $N=(1,0)$ supergravity in $D=6$.

In contrast to other applications of sigma models in field theory,
a characteristic property of gauged supergravity theories with
scalar fields is that they have potentials in their Lagrangians.
This is a consequence of the Noether procedure required to
establish local supersymmetry. Although the perturbative treatment
of string theories do not tend to produce gauged supergravity
theories, they do arise as low energy limits of M theory in
certain backgrounds in a nonperturbative framework. In this
context, it is natural to investigate the brane solutions of
gauged supergravity theories in diverse dimensions. In doing so,
the potentials mentioned above play an important role. Motivated
by this fact, we will elucidate the structure of a potential which
arises typically in gauged supergravity theories and we will
derive a general and simple formula for it. As an application, we
will apply this formula to the gauged $(1,0)$ supergravity in six
dimensions and derive an explicit formula for its potential. In
doing so, we also exhibit the relation between various
formulations of the gauged sigma models that exist in the
supergravity literature.

For completeness and in view of their possible applications in $D
<4$, we have also included a section on the gauged sigma models
with Wess-Zumino terms. This section is primarily based on
\citelow{hs}.

This paper contains the following sections:

\begin{enumerate}
  \item Introduction
  \item Minimal formulation
  \item Lifted formulation and coupling of fermions
  \item Gauging $K\subseteq G$
  \item Introducing a gauge invariant potential
  \item Adapted coordinates and $H$-gauge condition
  \item Introducing a gauged Wess-Zumino term
  \item Gauged sigma model on a bundle of frames
  \item Gauged sigma model on $G/H$
  \item The potential in $(1,0)$ supergravity $D=6$
  \item Appendix: Table of gauged supergravity theories in $D>5$
\end{enumerate}


\section{Minimal Formulation}


In its minimal form, a nonlinear sigma model is a theory of scalar
fields described by the Lagrangian

\be
{\cL}_\vf =-{1 \over 2 f^2} \sqrt{-\gamma}\gamma^{\m\n}
\p_\m \vf^\a \p_\n \vf^\b \,\, g_{\a\b} (\vf) \la{1.1}
\ee

where $\gamma_{\mu\nu}$ is the spacetime metric,
$g_{\a\b}$ is a function of the fields
(but not their derivatives), and $f$ is a coupling constant, which
will be set equal to 1 in the rest of the paper. This model can be
interpreted geometrically by saying that the fields
$\vf^\a$ are coordinate representatives of a map

\be
\vf: M \to N\ .
\ee

and that $g_{\a\b}$ is the metric on $N$ in the chosen coordinate
system. In field theory, and in particular in supergravity, $M$ is
interpreted as spacetime and $N$ as an internal space; in the
theory of extended objects, $M$ is the worldsheet and $N$ is
interpreted as spacetime. In the rest of the paper, we shall
assume that $M$ is flat Minkowskian spacetime, for simplicity.

It is usually desirable for physical reasons to assume that the
theory has global invariance under a symmetry group $G$.
Throughout this paper
$G$ will denote a Lie group, not necessarily compact; the Lie algebra of
$G$ will be denoted ${\cL}(G)$. We assume that in ${\cL}(G)$ there is
given an inner product, not necessarily $Ad(G)$-invariant, and
$\lbrace T_I \rbrace$, with $I=1,\ldots ,dim\,G $ will be an
orthogonal basis in
${\cL}(G)$. When the generators $T_I$ are represented by matrices, we
will assume that they are normalized so that
$Tr(T_IT_J)=-{1\over2}\d_{IJ}$. The structure constants $f_{IJ}{}^K$ are
defined by

\be
[T_I,T_J]=f_{IJ}{}^K T_K\ ; \la{1.3}
\ee

if the inner product in ${\cL}(G)$ is $Ad(G)$-invariant, then the
structure constants are totally antisymmetric (note that since the
metric in ${\cL}(G)$ is $\d_{IJ}$ the distinction between upper
and lower indices is immaterial).

In the following the components of all tensor fields on $N$ will
be referred to the natural bases $\lbrace \p_{\a}\rbrace$ and
$\lbrace dy^{\a}\rbrace$. The left action of $G$ on $N$ is
generated by vector fields $K_I=K_I{}^\a \p_\a$ which under Lie
brackets form an algebra anti-isomorphic to ${\cL}(G)$:

\be
{\cL}(G):\quad\quad
{\cL}_{K_I}K_J^\a = -f_{IJ}{}^L K_L{}^{\a}\ .
\la{1.4}
\ee

The reason for this minus sign is that conventionally ${\cL}(G)$
is defined as the algebra of left-invariant vector fields on $G$.
These vector fields generate the right action of $G$ on itself.
The left action of $G$ on itself is generated by the
right-invariant vector fields, whose algebra is anti-isomorphic to
${\cL}(G)$. Every left action of $G$ will be generated by vector
fields satisfying such an algebra.

For the action \eq{1.1} to be invariant, we assume that the
vectors
$K_I$ are Killing vectors for the metric $g$, that is, if ${\cL}_v$
denotes the Lie derivative along $v$,

\be
{\cL}_{K_I}g_{\a\b}=0\ .
\la{1.5}
\ee

If $\L $ is an element of ${\cL}(G)$, the infinitesimal variation
of the fields under {\it global} $G$ transformation is

\be
\d_\L\vf^\a=-\L^I K_I^\a(\vf)\ ,\quad\quad \p_\m \L^I=0\ .
\la{1.6}
\ee

In general, acting on any function of the fields, $\d_\L= \L^I
K_I^\a\p_\a $. Such variations satisfy an algebra isomorphic to
the abstract algebra ${\cL}(G)$: $
[\d_{\L_1},\d_{\L_2}]=\d_{[\L_1,\L_2]}$. Invariance of the action
based on the Lagrangian \eq{1.1} follows directly from using
\eq{1.5}.


\section{Lifted Formulation and Coupling of Fermions}


As we shall discuss later, in order to couple the system to
fermions it is sometimes necessary to use a different formulation
of the theory, where there are more fields than physical degrees
of freedom. Some of the fields (or functions thereof) are then
gauge degrees of freedom. This is completely analogous to what
happens in gravity, where the coupling to fermions requires the
use of the tetrad formalism.

The most general geometrical setup is to imagine a space $\bar N$
with a map

\be
\pi: \bar N \to N\ ,
\ee

which is surjective. In the new formulation the basic variables
will be fields $\bar\vf^{\bar\a}$, describing a map from spacetime
into $\bar N$. Given this map $\bar\vf$, one can construct in a
unique way a map
$\vf$ from spacetime into $N$ by composing $\bar\vf$ with the projection
$\pi$, and the Lagrangian must be constructed in such a way that it has
the same value for any two maps $\bar\vf$ that project onto the
same map
$\vf$.

This setup is unnecessarily general and in physical applications
it is usually assumed that the projection $\pi$ amounts to
factoring out the right action of some group $H$ acting on $\bar
N$. In the following we assume this to be the case.

Given a map $\vf:M\rightarrow N$, we say that a map

\be
\bar{\vf}: M \rightarrow \bar N \ ,
\ee

is a lift of $\vf$ if $\pi\bigl(\bar{\vf}(x)\bigr)=\vf(x)$. If
$\bar\vf$ is a lift of $\vf$, then also $\bar\vf^{\prime}$,
defined by
$\bar\vf^{\prime}(x)=\bigl(\bar\vf(x)\bigr)h(x)$ for some map
$h:M \rightarrow H$, is a lift of $\vf$. Therefore, the lifted nonlinear
sigma model has a nontrivial gauge group.

In general there are topological obstructions to the existence of
lifts \cite{rp1}. Here we will deal only with local properties and
we shall assume that lifts exist.

Let $f_{ab}{}^c$ be the structure constants of $H$. We have a
right action of
$H$ on $\bar N$, generated by vector fields $F_a=F_a^{\bar\a}\p_{\bar\a}$
whose algebra is isomorphic to ${\cL}(H)$:

\be
{\cL}(H):\quad\quad {\cL}_{F_a} F_b{}^{\c} = f_{ab}{}^c
F_c{}^{\c}\ .
\la{2.1}
\ee

Given an element $\eta=\eta^a T_a$ of ${\cL}(H)$, the
infinitesimal variation of the fields $\bar\vf$ is

\be
\d_\eta\bar\vf^{\bar\a}=\eta^a F_a^{\bar\a}(\bar\vf) \la{2.2}
\ee

and we have $ [\d_{\eta_1},\d_{\eta_2}]=\d_{[\eta_1,\eta_2]}$.

If there is a {\it global} invariance under $G$, it must be
realized also on the lifted fields with Killing vectors $\bar
K_I^{\bar\a}$ satisfying the same algebra \eq{1.4} as the fields
$K_I^\a$:

\be
\d_\L \bar\vf^{\bar\a}= -\L^I \bar K_I^{\bar\a}(\bar\vf)\ ,\quad\quad
\p_\m \L^I=0\ .
\la{2.4}
\ee

We must have $T\pi(\bar K_I)=K_I$ and this is possible if the
Killing vectors $\bar K_I$ are $H$-invariant, {\it i.e.}

\be
{\cal L}_{F_a}\bar K_I^{\bar\b}=0\ .
\la{2.5}
\ee

In order to rewrite the Lagrangian \eq{1.1} in terms of the lifted
fields, we need a new geometrical ingredient: a connection in the
bundle
$\pi:\bar N\to N$. By this we mean the following. The tangent space
$T_p\bar N$ at $p\in\bar N$ contains a subspace $V_p=ker T\pi$, called
the vertical subspace, which is tangent to the orbit of $H$. One
can take the vectors $F_a$ as a basis in $V_p$. There is, however,
no preferred choice of a complementary subspace in $T_p\bar N$. A
connection is precisely the assignment at each point $p$ of a
``horizontal'' subspace $H_p$ such that $H_p\oplus V_p=T_p\bar N$
and such that the distribution of these spaces is $H$-invariant:
for any
$h\in H$, $H_{ph}=H_p h$.

These horizontal spaces can be defined as the kernels of a
${\cL}(H)$-valued one-form $\o$ called the connection form, with
the properties that

\be
\o^{a}_{\bar\a}F_{b}{}^{\bar\a}= \d^{a}{}_{b}\ .
\la{2.6}
\ee

and

\be
{\cL}_{F_a}\o^b_{\bar\a}= -f_{ac}{}^b \o^c_{\bar\a}\ .
\la{2.7}
\ee

In addition, the connection is assumed to be $G$-invariant:

\be
{\cL}_{\bar K_I}\o^b_{\bar\a} =\, 0\ .
\la{2.8}
\ee

It follows from \eq{2.6} that the $G$ invariant tensors

\bea
V^{\bar\a}{}_{\bar\b} &=& F_{a}{}^{\bar\a}\o^{a}{}_{\bar\b}\ ,
\la{2.9a} \\
H^{\bar\a}{}_{\bar\b}
&=&\d^{\bar\a}{}_{\bar\b}-V^{\bar\a}{}_{\bar\b}\ ,
\la{2.9b}
\eea

are the vertical and horizontal projectors.

Next we define the covariant derivative of $\bar{\vf}$ by

\bea
D_{\mu}\bar{\vf}^{\bar{\a}} &=&
H^{\bar{\a}}{}_{\bar{\b}}\p_{\mu}\bar\vf^{\bar\b}
\nn\\
&=&\p_{\mu}\bar\vf^{\bar\a}-B_{\mu}^{a}F_{a}{}^{\bar\a}(\bar\vf)\ ,
\la{2.10}
\eea

where

\be
B_{\mu}^{a}=\p_{\mu}\bar\vf^{\bar\b}\o^{a}_{\bar\b}(\vf)
\la{2.11}
\ee

is a composite gauge potential which is inert under the global
left G transformations, and transforms as a gauge field under the
composite local right H transformations:

\bea
\d_\L B_\m^a &=& 0\ ,\nn\\
\delta_\eta B_\m^a &=& \p_\m B_\eta^a + f^a{}_{bc} B_\m^b \eta^c\ .
\la{2.12}
\eea

This result, together with \eq{2.7} implies that the covariant
derivative $D_\m{\bar\vf}^{\bar\a}$ transform as

\bea
\d_\L D_\m{\bar\vf}^{\bar\a} &=& -\L^I \p_{\bar\b}\bar K_I^{\bar\a}
D_\mu\bar\vf^{\bar\b}\ ,\la{2.13a}
\\
\d_\eta D_\mu\bar\vf^{\bar\a} &=&
\eta^a\p_{\bar\b}F_a^{\bar\a}D_\mu\bar\vf^{\bar\b}\ . \la{2.13b}
\eea

Let $\bar g =\bar g_{\bar\a\bar\b}\,d\bar y^{\bar\a} \otimes d\bar
y^{\bar\b}$ be a left $G$ and right $H$ invariant metric on $\bar
N$, such that $V\perp H$ and that given any vectors $(v,w)$ on
$N$, and the unique vectors $({\bar v}, {\bar w})$ on ${\bar N}$
which are horizontal and project to $(v,w)$, then the inner
product of
${\bar v}$ and ${\bar w}$ relative to ${\bar g}$ must be equal to the
inner product of $v$ with $w$ relative to $g$. The Lagrangian of
the lifted nonlinear sigma model can then be written as

\be
{\cL}_{\bar\vf}=-{1\over 2 }\,\bar g_{\bar\a\bar\b}(\bar\vf)
D^{\mu}\bar\vf^{\bar\a}D_{\mu}\bar\vf^{\bar\b}\ .
\la{2.14}
\ee

Because of its gauge invariance this Lagrangian depends really
only on
$\vf$ and it can be seen using \eq{2.10} that it coincides with the Lagrangian
\eq{1.1}.

We are now ready to couple fermions to the scalar fields. In the
standard way of doing this, one assumes that the fermions carry a
representation $\rho$ of the group $H$, so that when $\bar\vf$
undergoes \eq{2.2}, the fermion undergoes

\be
\d_\eta\psi=-\rho(\eta)\psi\ .
\la{3.1}
\ee

(The minus sign is necessary to ensure that these transformations
satisfy
$[\d_{\eta_1},\d_{\eta_2}]\psi=\d_{[\eta_1,\eta_2]}\psi$,
in accordance with \eq{2.2}.)

At each point $x$ in $M$ the fermion is given by an equivalence
class of pairs $({\bar\vf}(x),\psi(x))$ under $H$. Therefore the
fermion field can be thought of as a section of a vector bundle
associated to the pullback by $\vf$ of the principal $H$ bundle
${\bar N} \to N$. Note therefore that one cannot define what the
fermions are before having given a scalar field configuration.
This is in analogy with gravity where one cannot define what the
fermions are prior to having given a metric. Thus the
configurations space of scalars and fermions is not a product of
the scalar and fermion configuration spaces, but rather a fiber
bundle over the scalar configuration space.

In the action, a natural coupling between scalars and fermions
arises through the gauge covariant derivative of $\psi$, which is
defined by

\be
D_\mu\psi=\p_\mu\psi + B_\mu^a T_a \psi\ .
\la{3.2}
\ee

with $B_\mu$ defined as in \eq{2.11}. The fermionic kinetic term

\be
{\cL}_\psi= \ft12 {\bar\psi}\c^\m D_\m \psi\ ,
\ee

is manifestly $H$-gauge invariant.

To summarize, the total Lagrangian in the lifted formulation is
given by

\be
{\cL}_0 = -{1\over 2 }\,\bar g_{\bar\a\bar\b}(\bar\vf)
D^{\mu}\bar\vf^{\bar\a}D_{\mu}\bar\vf^{\bar\b}+
\ft12 {\bar\psi}\c^\m D_\m \psi\ ,
\la{Lzero}
\ee

where the covariant derivatives are defined in \eq{2.10} and
\eq{3.2}.


\section{Gauging $K \subseteq G$}


We consider now the gauging of any subgroup of $G$, denoted by
$K$, with generators $T_i\,(i=1,...,{\rm dim}\,K)$. In particular,
the group $K$ can be chosen to be $G$ or $H$. In addition to the
fields $\vf^\a$ we have a dynamical ${\cL}(G)$-valued gauge field
$A_\m =A_\m^i T_i$. Under an infinitesimal local
$K$-transformation we have

\bea
\d_\L {\bar\vf}^{\bar\a} &=& -\L^i(x){\bar K}_i^{\bar\a}\ ,
\la{4.1a}\\
\d_\L A_\m^i &=& \p_\m\L^i+g f^i{}_{jk}A_\m^j\L^k\ ,
\la{4.1}
\eea

where $g$ is the gauge coupling constant which will be set equal
to 1 in the rest of the paper. This definition is such that the
algebra \eq{1.4} is satisfied. A relative sign between the two
terms on the right hand side of \eq{4.1} can be absorbed by a
redefinition of $A$. Note also that $A_\m^i$ is $\eta$-invariant:

\be
\d_\eta A_\m^i = 0\ . \la{4.3c}
\ee

Next we define the $G$-covariant derivative of the lifted fields
as

\be
\nabla_\mu\bar\vf^{\bar\a}=
\p_\mu\bar\vf^{\bar\a}+A_\mu^i \bar K_i^{\bar\a} (\bar\vf)\ .
\la{4.4}
\ee

Upon using \eq{2.2}, \eq{2.5} and \eq{4.3c} one verifies that

\bea
\d_\L\nabla_\mu\bar\vf^{\bar\a} &=&
-\L^i (x) \p_{\bar\b}\bar K_i^{\bar\a} \nabla_\mu\bar\vf^{\bar\b}\
,
\la{4.5}\\
[0.2cm]
\d_\eta\nabla_\mu\bar\vf^{\bar\a} &=&
(\p_\mu\eta^a)F_a^{\bar\a}+
\eta^a\p_{\bar\b}F_a^{\bar\a}\nabla_\mu\bar\vf^{\bar\b}\ .
\la{4.7}
\eea

Using these transformation properties, and \eq{2.7}, one can check
that a composite $H$ gauge field defined by

\be
{\cB}_\mu^a=\nabla_\mu\bar\vf^{\bar\a}\o_{\bar\a}^a\ ,
\la{4.6}
\ee

transform under the local left $G$ and local right $H$
transformation as

\bea
\d_\L {\cB}_\mu^a &=& 0\ ,
\la{4.8}\\
[0.2cm]
\d_\eta {\cB}_\mu^a &=& \p_\mu\eta^a+f^a{}_{bc}{\cB}_\mu^b\eta^c\ ,
\la{2.12b}
\eea

where \eq{2.7} and \eq{2.6} have been used. It follows that if we
define the expression

\be
{\cD}_\mu\bar\vf^{\bar\a}=
\nabla_\mu\bar\vf^{\bar\a}-B_\mu^a F_a^{\bar\a}(\bar\vf)\ ,
\la{4.9}
\ee

it transforms as

\bea
\d_\L{\cal D}_\mu\bar\vf^{\bar\a} &=& -\L^i (x) \p_{\bar\b}\bar K_I^{\bar\a}
{\cD}_\mu\bar\vf^{\bar\b}\ ,
\la{4.11}\\
[0.2cm]
\d_\eta{\cal D}_\mu\bar\vf^{\bar\a} &=& \eta^a \p_{\bar\b}F_a^{\bar\a}
{\cal D}_\mu\bar\vf^{\bar\b}\ ,
\la{4.10}
\eea

so it deserves to be called the bi-covariant derivative of the
lifted field. One can now write the kinetic term in terms of the
lifted fields. In particular, the covariant derivative of the
fermions takes the form

\be
{\cD}_\mu\psi=\p_\mu\psi + {\cB}_\mu^a T_a \psi\ .
\la{3.2b}
\ee

Thus, a lifted gauged sigma model can be characterized by the
Lagrangian

\be
{\cL}= -{1\over 2 }\,\bar g_{\bar\a\bar\b}(\bar\vf)
{\cD}^{\mu}\bar\vf^{\bar\a} {\cD}_{\mu}\bar\vf^{\bar\b}
+{1\over 2} {\bar\psi}\c^\m {\cD}_\m \psi\ ,
\la{LG}
\ee

where the fermions $\psi$ carry a given representation of the
group $H$.


\section{ Introducing a Gauge Invariant Potential}


There is no unique way to construct a gauge invariant potential in
the context of bosonic sigma models. In the case of supersymmetric
sigma models, however, the requirement of supersymmetry is often
powerful enough to determine uniquely the form of the potential.
In particular, when one gauges the automorphism group of
supergravity theories which either contain scalar fields or are
coupled to matter multiplets which contain scalar fields, the
Noether procedure typically results in a potential. The important
building block for the potential arises in the process of
computing the supersymmetric variation of the gravitino kinetic
term

\be
{\cL}_{\psi_\m} = \ft12 e {\bar\psi}_\m \c^{\m\n\r} {\cD}_\n \psi_\r \ ,
\la{kt}
\ee

where $e$ is the determinant of the vielbein on $M$, and the
covariant derivative contains, in addition to the Lorentz
connection, a composite gauge field ${\cB}_\m^a T_a$ with $T_a$ in
the fundamental representation of the automorphism group
$H_{Aut}$.  In any supergravity theory the supersymmetric
variation of the gravitino must take the form

\be
\d \psi_\m = {\cD}_\m \e + \cdots\ , \la{susy}
\ee

where $\e(x)$ is the local supersymmetry parameter. Thus, in the
process of varying the kinetic term \eq{kt} under \eq{susy}, one
encounters the commutator term

\be
[{\cal D}_\mu,{\cal D}_\nu]\e ={\cG}_{\mu\nu}\e \ ,
\la{com}
\ee

where ${\cG}_{\mu\nu}={\cG}_{\mu\nu}^a T_a$ is the
${\cL}(H)$ valued curvature tensor of the composite connection:

\be
{\cG}_{\mu\nu}^a :=\p_\mu {\cB}_\nu^a-\p_\nu {\cB}_\mu^a
+f_{bc}{}^a {\cB}_\mu^b {\cB}_\nu^c\ ,
\la{gmn}
\ee

and $T^a$ is in the fundamental representation of $H_{Aut}$. From
the definition \eq{4.6} it is straightforward to compute:

\be
{\cG}_{\mu\nu}^a  = F_{\mu\nu}^i C_i{}^a \ ,
\la{gmn2}
\ee

where $F_{\mu\nu}=F_{\mu\nu}^i T_i$ is the ${\cL}(K)$-valued
curvature of $A_\mu$ and

\be
C_i{}^a = \bar K_i^{\bar\a}\o_{\bar\a}^a\ .\
\la{cf}
\ee

Thus, from \eq{kt}, \eq{susy} and \eq{com} we see that a term of
the form

\be
e {\bar\psi}_\m \c^{\m\n\r}T_a\epsilon\,F_{\n\r}^i C_i^a\ ,
\ee

has to be cancelled by supersymmetric variation of other terms.
The time tested Noether procedure to establish local supersymmetry
then quickly leads to Yukawa couplings involving the
$C$-function and a potential of the form (see, for example,
\citelow{ns} for details of how exactly this works)

\be
{\cL}_C = - e^{-\vf} \,{\rm tr}\, C_{i} C^{i} \ ,
\la{pot}
\ee

where $\vf$ is the dilaton coming from the tensor multiplet and

\be
C_i := C_i^a T_a\ .
\ee

Using \eq{1.4}, \eq{2.8}, \eq{2.5} and \eq{2.7} it is easy to
verify that

\bea
{\cL}_{K_i} C_j^a &=& f_{ij}{}^k C_k^a\ ,
\nn\\ [0.2cm]
{\cL}_{F_a} C_i^b &=& f_{ca}{}^b C_i^c\ .
\eea

Therefore the Lagrangian \eq{pot} is local left $G$ and local
right $H$ invariant.

It should be emphasized that although the Noether procedure
mentioned above primarily arises in the context of sigma model
manifolds that are coset spaces, we can nonetheless introduce the
potential \eq{pot} for general sigma model manifolds, as it has
all the right properties. This will be done in what follows, until
we come back to the application of the general formalism to
specific examples. Without loss of generality, we will also
continue to consider Minkowskian spacetimes.


\section{Adapted Coordinates and $H$-Gauge Condition }


As we said in section 2, the choice of coordinates on $N$ is
completely arbitrary: different choices amount to field
redefinitions and thus lead to the same physical results.
Similarly, in a lifted formulation the coordinates on $\bar N$ are
also completely arbitrary. In practice it is sometimes convenient
to use this freedom to choose a coordinate system adapted to the
bundle structure. This means choosing locally a diffeomorphism of
$\bar N$ to  $N\times H$, and using coordinates $y^\a$ on $N$ and
$y^{\hat\a}$ on $H$ as coordinates on $\bar N$. We will then divide
the fields as

\be
{\bar\vf}^{\bar\a}=(\vf^\a,\vf^{\hat\a})\ .
\ee

where, obviously, $\bar\varphi^\alpha=\varphi^\alpha$ and
$\vf^{\hat\a}(x)$ are the coordinate components of an abstract
$H$-valued field $h(x)$. In this coordinate system the fundamental vector
fields have components

\be
(F_a^\a,F_a^{\hat \a})=(0,L_a^{\hat \a})\ . \la{ff}
\ee

where $L$ are the left-invariant vector fields on $H$. The
connection at a point with coordinates $(y^\a, h)$ can be
represented as

\be
\o=h^{-1}B(y^\a) h+h^{-1}dh\ ,
\la{2.15}
\ee

$B^b(y^\a)$ being the local gauge potential on $N$ and $h^{-1}dh$ the
left-invariant Maurer-Cartan form on $H$. Thus $\o$ has components

\bea
\o^a_\a &=& Ad\,(h^{-1})^a{}_b B^b_\a(y)\ ,  \nn\\
\o^a_{\hat\a} &=& L^a_{\hat a}(y^{\hat\a})\ . \la{2.15b}
\eea

Using these components and the fact that

\be
L_a^{\hat\b}\p_{\hat\b}Ad\,(h^{-1})^b{}_c= -f_{ad}{}^b
Ad\,(h^{-1})^d{}_c\ ,
\ee

one can verify separately the $\a$ and ${\hat\a}$ components of
\eq{2.7}.

Since $\bar K_I$ projects onto $K_I$, there is a unique generator
$v_I=v_I^a T_a$ of $H$ such that

\be
\bar K_I=K_I+v_I^a F_a \ . \la{kk}
\ee

In adapted coordinates, the components of $\bar K_I$ are therefore
given by ${\bar K}_I^{\bar \a}= ( K_I^\a, v_I^a L_a^{\hat\a})$.
The vector
$v_I^a$ depends both on $y^\a$ and $y^{\hat \a}$, as one gets from
\eq{2.5}

\be
L_a^{\hat\a}\p_{\hat\a}v_I^c=-f_{ab}{}^c v_I^b \ .
\la{dv}
\ee

Using this relation one finds that the $\hat\a$ component of
\eq{2.8} is identically satisfied, and that the $\a$ component
gives

\be
{\cL}_{K_I}\o^b_\a =(\p_\a v_I^b+f^b{}_{ca}\,\o^c_\a v_I^a)\ .
\la{2.16}
\ee

For $h=1$ this gives the familiar statement that a connection is
invariant if the Lie derivative of the gauge potential is an
infinitesimal gauge transformation. The advantage of using adapted
coordinates is that they provide a clean separation of the gauge
degrees of freedom (the $\vf^{\hat\a}$) from the physical degrees
of freedom (the $\vf^\a$). Once the gauge degrees of freedom have
been thus isolated, one can choose a gauge by simply fixing the
functions $\vf^{\hat\a}$. For example, we may choose

\be
\vf^{\hat\a}= \vf^{\hat a}_0 \ ,
\la{hg}
\ee

where ${\vf}^{\hat \a}_0$ is a constant. In this gauge, the sum of
the Lagrangian \eq{LG} and \eq{pot}, where a subgroup $K$ of the
full isometry group $G$ has been gauged, takes the form

\be
{\cL}= -{1\over 2 }\,g_{\a\b}(\vf) {\cD}^{\mu}\vf^{\a}
{\cD}_{\mu}\vf^{\b} +{1\over 2} {\bar\psi}\c^\m {\cD}_\m \psi\ -
{\rm tr}\, C_i C^i\ ,
\la{LG2}
\ee

where we have suppressed the tensor multiplet dilaton and

\bea
{\cD}_\m \vf^\a  &=& \p_\m \vf^\a + A_\m^i K_i^\a\ ,\la{def1}\\
[0.25cm] C_i^a &=& K_i^\a B_\a^a + v_i^a\ ,
\la{def2}
\eea

and ${\cB}_\m^a$ occurring in the covariant derivative
\eq{3.2b} takes the form

\be
{\cB}_\m^a = {\cD}_\m \vf^\a B_\a^a\ T_a + A_\m^i v_i^a\ T_a\ .
\la{cd1}
\ee

Note that (62) is the $G$-covariant derivative of the (unlifted)
field $\varphi$ and the first term in (61) is just the gauged
version of (1). We recall that $v_i^a (\vf)$ is a function of the
scalars which is to be determined from \eq{kk}. In section 9 we
will derive a general and simple formula for the $C$-function in
the case
$N=G/H$, without having to compute the exact form of $v_i^a$.


\section{Introducing a Gauged Wess-Zumino Term}


In addition to the kinetic term and a potential, nonlinear sigma
models may also contain higher derivative terms, or terms that are
linear in the time derivative. A term of the latter kind is of
particular interest and is known as the Wess-Zumino term.
Nonlinear sigma models with Wess-Zumino terms are known as the
Wess-Zumino-Witten (WZW) models. There exists a vast literature on
this subject. Here we shall only review a general action formula
valid in arbitrary dimensional spacetime $M$ and for scalar fields
taking their values in an arbitrary riemannian manifold $N$.

Let $M$ be $(p+1)$-dimensional, and let us define the following
forms on $N$:

\bea
&& b = \ft1{(p+1)!}\, d\vf^{\a_1}\cdots
d\vf^{\a_{p+1}}\,b_{\a_1\cdots\a_{p+1}}\ ,
\quad\quad H=db\ ,
\nn\\ [+0.2cm]
&& v_{i_1\cdots i_k}^{(r)} = \ft1{r!}\,d\vf^{\a_1} \cdots d\vf^{\a_r}\,
v_{\a_1\cdots \a_r, i_1\cdots i_k}=v_{(i_1\cdots i_k)}^{(r)} \ ,
\nn\\[+0.2cm]
&& r=p,p-2,...,\vare\ , \quad\quad  2k+r=p+2\ ,
\nn\\[+0.2cm]
&& \vare=0 \ {\rm for\ even}\ p\ ,
\quad\quad \vare=1\ {\rm for\ odd}\ p\ .
\eea

As in the previous sections, we assume that $N$ has the isometries
generating the group $G$, and we gauge the $K$ subgroup of $G$
generated by the Killing vectors $K_i^\a$. We shall work in the
adapted coordinate system described in the previous section.

Let $M$ be the boundary of a $(p+2)$-dimensional manifold $Y$, and
let us define the following covariant pull-backs to $Y$ ( For the
purposes of this section only, we will adhere to the convention of
\citelow{hs} for the covariant derivative according to which the
replacement $A \ra -A$ is to be made in \eq{def1}):

\bea
&& \tH^{(p+2)} = \ft1{(p+2)!}\,dx^{\m_1} \cdots dx^{\m_{p+2}}\,
{\cD}_{\m_1}\vf^{\a_1}\cdots {\cD}_{\m_{p+2}}\vf^{\a_{p+2}}\,
H_{\a_1\cdots\a_{p+2}}\ ,
\nn\\ [+0.2cm]
&& \tv_{i_1\cdots i_k}^{(r)} = \ft1{r!}\, dx^{\m_1} \cdots dx^{\m_{r}}\,
{\cD}_{\m_1}\vf^{\a_1}\cdots {\cD}_{\m_{r}}\vf^{\a_{r}}\,
v_{\a_1\cdots \a_r, i_1\cdots i_k}\ .
\eea

Provided that $\tH$ and $\tv$ satisfy certain conditions (see
below) the gauged WZW action can be written as an integral over
$Y$ as follows \cite{hs}

\bea
S_{GWZ} & = &\int_Y \left( \tH^{(p+2)} + \tv_i^{(p)} F^i +
\tv_{ij}^{(p-2)}F^i F^j+\cdots +
\tv_{i_1\cdots i_N}^{(\vare)} F^{i_1}\cdots F^{i_N} \right)\ ,\nn\\
& \equiv  & \int_Y {\cL}_{(p+2)}\ ,\quad\quad N = \ft12 (p+2-\vare)\ .
\la{gwz}
\eea

Each term in this action is separately gauge invariant provided
that

\bea
&& {\cL}_{K_i} H = 0\ , \nn\\ [+0.2cm]
&& {\cL}_{K_j} v_{i_1\cdots i_k}^{(r)} -k\, f_{j(i_1}{}^\ell \,
v_{i_2\cdots i_k)\ell}^{(r)}=0\ .\la{c12}
\eea

The set of forms $v^{(r)}$ are needed, however, so that the
Lagrangian in \eq{gwz} is closed. This property makes it possible
to write the action on the boundary of $Y$. Indeed, using the
following identity, which is valid for any covariantly pulled-back
form $\tO$,

\be
d\wt{\O}=\wt{d\O}-F^j \left( \wt {{i_{K_j}}\O} \right) + A^j
\left(
\wt{{\cL}_{K_j}\O}\right)\ ,
\ee

and using \eq{c12}, one can show that

\be
d{\cL}_{p+2}=0\ ,
\ee

provided that the following additional conditions are satisfied
\cite{hs}

\bea
i_k H^{(p+2)} &=& d v_k^{(p)}\ ,
\nn\\[+0.2cm]
i_{(j} v_{i)}^{(p)} &=& dv_{ij}^{(p-2)}\ ,
\nn\\[+0.2cm]
i_{(k} v_{ij)}^{(p-2)} &=& dv_{ijk}^{(p-4)}\ ,\nn\\
&\vdots& \nn\\
i_{(i_1} v_{i_2...i_N)}^{(\vare+2)} &=& dv_{i_1...i_N}^{(\vare)}
\ ,
\eea

where we have used the notation $i_{K_n} \equiv i_n$. Note that if
$H^{(p+2)}$ satisfies the property $i_k H^{(p+2)}=0$, then the
$v$-forms would not be necessary for gauging of the WZW model; one simply
makes the replacement $\p_\m\vf \ra {\cD}_\m \vf$ in that case.

In order to write the action as an integral over the $(p+1)$
dimensional manifold $M=\p Y$, we need the variation of
${\cL}_{p+2}$ with respect to $F^i$, which will be denoted by
$K_i$.

\be
K_i={\d{\cL}\over \d F^i}\ .
\ee

Then, as shown in \citelow{hs}, the action on $M$ can be written
as

\be
S_{GWZ} = \int_M \int_0^1 dt\, A^iK_i(tA)\ ,
\ee

where it is understood that the substitution $ A\rightarrow tA $
is to be made everywhere the gauge potential $A$ occurs in the
functional $K$.

For $p=2$, for example, the gauged Wess-Zumino action takes the
form \citelow{hs}

\bea
S_{GWZ}=&& \int_M \Bigl( b^{(3)} + A^iv_i^{(2)} +\ft12 A^iA^j
v_{ij}^{(1)} -\ft16 A^iA^jA^k v_{ijk}^{(0)}\nn\\
&& \quad\quad
+v_{ij}^{(0)} ( A^idA^j +\ft13 A^i f_{kl}{}^j A^k A^l ) \Bigr)\ ,
\eea

where we have the earlier definitions

\be
i_{k} H^{(3)} = dv_k^{(2)}\ ,
\quad\quad i_{(k} v_{j)}^{(2)}=v_{jk}^{(0)}\ ,
\ee

as well as new ones defined in terms of these as

\be
dv_{ij}^{(0)} := v_{ij}^{(1)}\ ,\quad\quad i_{(k} v_{ij)}^{(1)} :=
v_{ijk}^{(0)}\ .
\ee

We conclude this section by noting that the gauged WZW model in
arbitrary dimension with fundamental gauge fields on $M$ is
closely related to a general gauged sigma model studied in the
context of bosonic $p$-branes \cite{dds,ps} where the gauge fields
are not fundamental vector fields on the worldvolume $M$, but
rather they are the target space background fields. The issue of
gauge anomalies acquires different significance in these two
cases.


\section{Gauged Sigma Model on a Bundle of Frames}


An example of a lifted sigma model encountered in supergravity is
when
$N$ is any riemannian manifold and $\bar N=LN$ is the bundle of
linear frames of $N$. In this case $H=GL(n)$, where $n$ is the
dimension of $N$. The adapted coordinates in this case consist of
coordinates on $N$ and a matrix-valued field $e^\a{}_\b$
representing a general basis on $N$ (the index $\hat\a$ in this
case consists of the pair of indices $(\a,\b)$.) The gauge
potential $B_\c^a$ is given by the components of the linear
connection $\C_\c{}^\a{}_\b$, where $\c$ is the form index and
$(\a,\b)$ are the Lie algebra index. We take the connection to be
the Levi-Civita connection of the metric $g_{\a\b}$. Since $g$ is
assumed to be invariant, also the corresponding linear connection
is invariant. Under an infinitesimal isometry generated by $K_I$,
the Levi-Civita connection transforms as

\be
{\cL}_{K_I}\C_\c{}^\a{}_\b = -(\p_\c\p_\b K_I^\a +\C_\c{}^\a{}_\d
\p_\b K_I^\d -\C_\c{}^\d{}_\b \p_\d K_I^\a)\ .
\la{2.18}
\ee

which is just \eq{2.16} in the gauge $h=1$ and with
$v_I{}^\b{}_\a=\p_\a K_I^\b$.

Consider the gauging of the $K$ subgroup of the full isometry
group $G$. Under an infinitesimal isometry \eq{4.1a} a linear
basis $e^\a{}_\b$ transforms as

\be
\d_\Lambda e^\a{}_\b=-\L^i\,\p_\c K_i^\a\,e^\c{}_\b\ .
\la{4.12}
\ee

Therefore $\bar K_i{}^\a{}_\b=\p_\c K_i^\a e^\c{}_\b$. From
\eq{def1} we thus have

\be
{\cD}_\mu e^\a{}_\b=
\p_\mu e^\a{}_\b+A_\mu^i\, \p_\c K_i^\a\, e^\c{}_\b\ .
\la{4.13}
\ee

Under \eq{4.1a} we find

\be
\d_\L{\cD}_\mu e^\a{}_\b=
-\L^i\,\p_\c K_i^\a\,{\cD}_\mu e^\c{}_\b +\L^i{\cD}_\mu\vf^\d\,
\p_\d\p_\c K_i^\a\, e^\c{}_\b\ .
\la{4.14}
\ee

which is in accordance with \eq{4.5}. For the composite gauge
potential in adapted coordinates one finds

\be
{\cB}_\mu{}^\a{}_\b= e^{-1\a}{}_\d{\cD}_\mu\vf^\c
\C_\c{}^\d{}_\phi e^\phi{}_\b
+e^{-1\a}{}_\d {\cD}_\mu e^\d{}_\b\ .
\la{4.15}
\ee

The invariance of this potential under $G$ follows by using
\eq{2.18}, whereas under

\be
\d_\eta\vf^\a=0\ , \quad\quad
\d_\eta e^\a{}_\b=\eta^\a{}_\c e^\c{}_\b\ ,
\la{4.16}
\ee

one finds again \eq{2.12b}.

The fermions carry a representation of the group $GL(n)$ and under
\eq{4.16} transform as

\be
\d_\eta \psi^\a=-\eta^\a{}_\b \psi^\b\ .
\ee

Under an infinitesimal isometry \eq{4.1a} with the attendant
transformation \eq{4.12} of the linear frames, one finds

\be
\d_\L \psi^\a=-\L^i \p_\b K_i^\a \psi^\b\ .
\ee

Eq. \eq{3.2b} becomes

\be
{\cD}_\mu \psi^\a=\p_\mu \psi^\a +{\cD}_\mu\vf^\c e^{-1\a}{}_\d
\C_\c{}^\d{}_\phi e^\phi{}_\lambda \psi^\lambda
+e^{-1\a}{}_\c{\cD}_\mu e^\c{}_\d \psi^\d \ .
\ee

Upon using the $H$ gauge freedom one can choose
$e^\a{}_\b=\d^\a_\b$, in which case

\be
{\cD}_\mu \psi^\a=\p_\mu \psi^\a +{\cD}_\mu\vf^\c
\C_\c{}^\a{}_\b \psi^\b -A_\mu^i \p_\b K_i^\a \psi^\b\ .
\ee

Taking into account obvious notational differences (a redefinition
of $\L$ by a sign), this corresponds to the formula given in
\citelow{bagger}.


\section{Gauged Sigma Models on $G/H$}


The most frequently encountered sigma models are based on coset
spaces. Let us assume therefore that $N=G/H$, where the coset
space $G/H$ is reductive, {\it i.e.} there exists an
$Ad(H)$-invariant subspace ${\cP}$ of ${\cL}(G)$ such
that

\be
{\cL}(G)={\cL}(H) \oplus {\cP}\ . \la{1.9}
\ee

The space ${\cP}$ can be identified with the tangent space to
$G/H$ at the coset $eH$. Note that if the basis is chosen in such a way
that $\lbrace T_a\rbrace$ with $a=1,\ldots,dim\,H$ is a basis in
${\cL}(H)$ and $\lbrace T_{r}\rbrace$ with $r=1,\ldots,dim\,G/H$)
is a basis in ${\cP}$, then

\be
f_{ab}{}^{r}=0\ \ \ ;\ \ \ f_{a r}{}^b =0\ .
\la{1.10}
\ee

The group $G$ acts on $G/H$ from the left by
$g(g^{\prime}H)=(gg^{\prime})H$. On the group we have
left-invariant and right-invariant vector fields $L_I$ and $R_I$.
They are chosen to agree at the identity:
$L_I(e)=R_I(e)$, and they commute: $[L_I,R_I]=0$. The vector
fields $K_I$ generating the left action of $G$ are the projections
of the right-invariant vector fields $R_I$ under the map
$g\to gH$. We assume that the restriction to ${\cal P}$ of the inner
product in ${\cL}(G)$ is $Ad(H)$-invariant; via standard theorems,
this gives rise to a $G$-invariant metric
$g=g_{\a\b}\,dy^{\a}\otimes dy^{\b}$ on $G/H$.

In the lifted formulation, we have $\bar N=G$, $\bar K_I=R_I$ (the
right-invariant vector fields on $G$) and $F_a=L_a$ (the
left-invariant vector fields on $H$). For the invariant connection
we take the ${\cal L}(H)$-component of the left-invariant
Maurer-Cartan form $g^{-1}\p_{\bar\alpha} g$ on $G$. This example
illustrates the reason why the groups are chosen to act as they
were. Traditionally one chooses to work with right cosets $gH$.
This fixes the action of $H$ on
$G$ to be from the right. The remaining action of $G$ on the coset
space is from the left. It arises from the action of $G$ on itself
from the left.

We shall now review a well-known way of writing sigma models in
terms of matrices, and recast the earlier results in this
formalism. Traditionally one works in a gauge-fixed version of the
lifted formalism, the gauge fixing being given by a locally
defined section $L: G/H \to G$. This section is just a choice of a
coset representative for each coset. In addition, as usual when
working with groups, it is very convenient to use matrix
representations, so we also write $L(y)$ for the matrix
representing the abstract group element $L(y)$. Under the action
of a group element $g$, $y\to y'$ and

\be
L(y')=g L(y)h^{-1}\ , \la{gLh}
\ee

where $h=h(g,y)$ is a compensating gauge transformation that
restores the chosen gauge. Infinitesimally, if $g=1+\Lambda$, we
can write $h=1+v$, where $v=v(y,\Lambda)$ is the matrix
representing the Lie algebra element $v$ that was defined in
\eq{kk}. Inserting in \eq{gLh} one gets the formula

\be
K_i^\a \p_\a L = T_i L-L v_i^a T_a\ , \la{KL}
\ee

which is just a matrix way of rewriting \eq{kk} (the right
invariant vector field $R_I$ at $L(y)$ is represented by the
matrix $T_I L(y)$ and so on).

The pull-back the Maurer-Cartan form by the section $L$ can be
decomposed as

\be
L^{-1}\p_\a L=V_\a^r T_r+ B_\a^a T_a\ ,\la{mc2}
\ee

where $V_\a^r$ is the vielbein and $B_\a^a$ is a gauge potential
on $G/H$. It is also convenient to define

\be
L^{-1}\p_\m L=P_\m^r T_r+ B_\m^a T_a\ ,\la{mc1}
\ee

where

\be
P_\m^r = \p_\m \vf^\a V_\a^r\ ,\quad\quad B_\m^a=\p_\m\vf^\a
B_\a^a \ .
\ee

It is easy to show that $P_\m^r$ transforms covariantly and
$B_\m^a$ as a gauge field under the composite local $H$-transformations.
Indeed, $B_\m^a$ coincides with \eq{2.11} in adapted coordinates
(see \eq{2.15b}).

The ungauged sigma model Lagrangian \eq{Lzero} can be written as

\be
{\cL}_0 = \ft12 P_{\m r} P^{\m r} + {\bar \psi} \c^\m
\left (\p_\m + B_\m^a T_a \right)\psi \ .
\ee

Upon the gauging of a subgroup $K$ of $G$, the decomposition
\eq{mc1} has to be modified as follows

\be
L^{-1}\left( \p_\m+ A_\m^i T_i\right) L = {\cP}_\m^r T_r+
{\cB}_\m^a T_a\ .
\la{mc3}
\ee

We will now show that the $H$-connection ${\cB}$ takes the form
given earlier in \eq{cd1} and that the quantity $P_\m^r$ can also
be represented in terms of the covariant derivative of the
scalars. To this end, we multiply \eq{KL} with $L^{-1}$ from the
left and use \eq{mc2} to obtain

\bea
K_i^\a V_\a^r= \left(L^{-1}T_i L\right)^r \ ,
\la{kv}\\[+0.2cm]
K_i^\a \o_\a^a = \left(L^{-1}T_i L\right)^a-v_i^a\ .\la{ko}
\eea

Using these relations in \eq{mc3}, we find

\bea
{\cP}_\m^r &=& {\cD}_\m\vf^\a V_\a^r\ ,
\nn\\[+0.2cm]
{\cB}_\m^a &=& {\cD}_\m \vf^\a B_\a^a + A_\m^i v_i^a\ .
\la{rels}
\eea

As a by product, we find that the expression for the $C$-function
given in \eq{def2} can now be written as

\be
C_i^a = \left(L^{-1}T_i L\right)^a \ .
\la{cf2}
\ee

In summary, the gauge invariant sigma model Lagrangian \eq{LG2}
can be written as

\be
{\cL} = \ft12 {\cP}_{\m r} {\cP}^{\m r} +{\bar \psi}
\c^\m \left(\p_\m + {\cB}_\m^a  T_a \right) \psi - {\rm tr}\, C_i C^i \ .
\ee

As an application of the formula \eq{cf2}, in the next section we
will compute the potential that arises in the gauged $(1,0)$
supergravity in six dimensions. The formula \eq{cf2} can readily
be applied also to a class of supergravity theories where
$N=G/H$ and the dimension of the gauge group $K$ equals the dimension
of the defining representation of $G$. In these cases, the
generator $T_i$ occurring in \eq{cf2} becomes a structure constant
of the group $K$ \cite{ss2,gst,cas}.


\section{The Potential in $(1,0)$ Supergravity in $D=6$ }


The $n$ copies of the hypermultiplets in this theory parametrize a
noncompact quaternionic coset manifold \cite{ns}. A generic
example of such a manifold is

\be
{G\over H} = {Sp(n,1)\over Sp(n)\times Sp(1)}\ , \la{coset}
\ee

where $Sp(1)$ is the automorphism group of the supersymmetry
algebra. Thus the supersymmetry parameter $\e^A(x)$ carries the
$Sp(1)$ doublet index $A=1,2$. The group $K=Sp(1)\times Sp(1)
\subset SP(n,1)$ has been gauged in \citelow{ns}. The group $K'$
in Table 1 refers to $Sp(n)$ in this example. Let the index
$a'=1,...,2n$ label the fundamental representation of this group.
The index $i=1,..., {\rm dim}\ K$ splits into the symmetric pairs
$(AB)$ and $(a'b')$, and we have the $C$-function matrices

\bea
&& C_i^a \ \ \ra\ \ \left(\,C^a_{AB}\ ,\ C^{a}_{a'b'}\, \right)
\nn\\[+0.2cm]
&&
a=1,2,3, \quad\quad A=1,2, \quad\quad a'=1,...,2n\ .
\eea

From \eq{cf2} we have

\be
C^a_{AB} = \left(L^{-1} T_{AB} L\right)^a\ ,\quad\quad C^a_{a'b'}
= \left(L^{-1} T_{a'b'} L\right)^a \ ,
\la{ca}
\ee

where $T_{AB}$ and $T_{a'b'}$ are the generators of $Sp(1)$ and
$Sp(n)$, respectively. To compute the explicit form of these functions,
let us choose the standard representation of the coset \eq{coset}
as follows

\be
L= {\rm exp}\ \left(\begin{tabular}{cc}
  0 & $\vf_{a'}{}^A$ \\
  $(\vf^{b'}{}_B)^T$ & 0 \\ \end{tabular}\right)\ ,
\ee

where the scalar fields satisfy the following conditions

\be
\vf^{b'}{}_B = \left(\vf_{b'}{}^B\right)^* = \O^{b'a'}\,
\vf_{a'}{}^A\,\vare_{AB}
\ee

The $\O$ and $\vare$-tensors are constant antisymmetric invariant
tensors of $Sp(n)$ and $Sp(1)$, respectively. Using matrix
notation, we have

\be
L=\left(\begin{tabular}{cc} cosh $\sqrt{\vf\vf^\dagger}$ &
$\vf\ {{\rm sinh}\ \sqrt {\vf^\dagger\vf}\over \sqrt{\vf^\dagger\vf}}$ \\
$\vf^\dagger\
{{\rm sinh}\ \sqrt {\vf\vf^\dagger}\over \sqrt{\vf\vf^\dagger}}$&
cosh $\sqrt{\vf^\dagger \vf}$ \\ \end{tabular}\right)\ ,
\la{L2}
\ee

where $\vf$ represents an $n\times 2$ matrix. Note that
$(\vf^\dagger\vf)_A{}^B=\vf_A{}^{a'} \vf_{a'}{}^B= \ft12 \vf^2 \d_A^B$
where $\vf^2 := {\rm tr}\ \vf^\dagger\vf$. We can map a pair of
symmetric $Sp(1)$ indices to an $Sp(1)$ vector index through the
relation $V^a =\ft12 (\s^a)_A{}^B V_B{}^A$. Let us also recall the
explicit form of the defining representations of $Sp(1)$ and
$Sp(n)$:

\be
(T_{AB})_C{}^D = \vare_{CB}\d_A^D + \vare_{CA}\d_B^D\ ,
\quad\quad\quad
(T_{a'b'})_{c'}{}^{d'} = \O_{c'b'} \d_{a'}^{d'} +
\O_{c'a'}\d_{b'}^{d'}\ . \la{dr}
\ee

From \eq{ca}, \eq{L2} and \eq{dr}, we find

\bea
C^a_{AB} & = & {\rm cosh}^2\ \sqrt{\ft{\vf^2}{2}}\ \s^a_{AB}\
\nn\\[0.3cm]
C^a_{a'b'} & = &{2\over \vf^2}\,{\rm sinh}^2\
\sqrt{\ft{\vf^2}{2}}\ \left( \vf \s^a \vf^\dagger \right)_{a'b'}\ .
\eea

The total potential in the gauged $(1,0)$ supergravity in $D=6$
constructed in \citelow{ns} is the sum of the squares of these
$C$-functions multiplied by an exponent of the dilaton field which
comes from the tensor multiplet. The $C$-functions were also
computed in \citelow{ns} but for a different choice of the coset
representative $L$. This corresponds to a redefinition of the
scalar fields and therefore the physical content is the same.

\section*{Acknowledments}

It is a pleasure to dedicate this paper to Richard Arnowitt. We
thank P.S. Howe and S. Randjbar-Daemi for helpful discussions.
This research has been supported in part by NSF Grant PHY-9722090
and the European Grant ERBFMRXCT960090.

\clearpage
\t1
\clearpage


\section*{References}


\ed
